\documentclass[a4paper,11pt]{article}
\pdfoutput=1 % if your are submitting a pdflatex (i.e. if you have
\usepackage{jheppub} % for details on the use of the package, please
\usepackage[T1]{fontenc} % if needed

\def\ba{\begin{eqnarray}}
\def\ea{\end{eqnarray}}

\title{\boldmath Entropy Linear Response Theory with Non-Markovian Bath }

\author{Yu Chen $^a$}

\affiliation[a]{Graduate School of China Academy of Engineering Physics , P. R. China}
\emailAdd{ychen@gscaep.ac.cn}

\abstract{We developed a perturbative calculation for entropy dynamics considering a sudden coupling between a system and a bath. The theory we developed can work in general environment without Markovian approximation. A perturbative formula is given for bosonic environment and fermionic environment, respectively. We find the R\'{e}nyi entropy response is only related to the spectral functions of the system and the environment, together with a specific statistical kernel distribution function. We find a $t^2$ growth/decay in the short time limit and a $t$ linear growth/decay in longer time scale for second R\'{e}nyi entropy. A non-monotonic behavior of R\'{e}nyi entropy for fermionic systems is found to be quite general when the environment's temperature is lower. A Fourier's law in heat transport is obtained when two systems' temperature are close to each other. A consistency check is made for Sachdev-Ye-Kitaev model coupling to free fermions, a Page curve alike dynamics is found in a process dual to black hole evaporation. An oscillation of entanglement entropy is found for a gapped spectrum of environment.}


\begin{document} 
\maketitle

\section{Introduction}

Linear response theory (LRT) is essencial for studying quantum matters, it lays the foundation for physical interpretations of the dynamical response after a perturbative probe to the system\cite{Mahan}. LRT is widely used in all kinds of condensed matter experiments, cold atoms experiments, such as APRES, conductivity, neutron scattering, etc. In original linear response theory, the dynamics of physical observables are considered in response to external sources. These perturbations are hermitian couplings to the systems. However, when we consider a more general case for dissipative couplings, a new response theory is needed and established as a Non-hermitian linear response theory (NHLRT) \cite{Hui20}. A relation between dissipative dynamics \cite{DissBH} and the correlation function in initial equilibrium state can be then established with the help of NHLRT. 

However, previous response theories are mostly focused on physical observables, or for dissipative couplings with Markovian approximations, entropies response theories for a general environment is still not fully established. In this paper, we are going to establish a general perturbative response theory for entropy with a generic bath. Nowadays, thermal entropies as well as entanglement entropies become increasingly important in studying eigenstate thermalization\cite{Deutsch91,Srednicki94,Cardy04,Grover18}, information scrambling\cite{Qi16,Ruihua17} and quantum gravity\cite{TBook}. It is in the core of holographic duality\cite{Maldacena98} which bridges the conformal field theory and the gravity through Ryu-Takayanagi formula\cite{RT06,T07,Wall15}. In addition to these theoretical interests, new technical advances are also achieved recently in experimental measurement of R\'{e}nyi entropy (RE) \cite{Greiner15, Greiner16,Xinhua17}. All these advances in both the theoretical and experimental studies make a general response theory for entropy a necessity.

Following the spirit of taking the coupling between the environment and the system as a perturbation, we can establish a linear response theory of R\'{e}nyi entropy for a general system with a sudden coupling to a general environment. Here we assume a general non-Markovian bath. The term non-Markovian means the temperature and spectrum of the environment can be arbitrary. Now we summarize our results as follows. We find that to the lowest order of coupling strength, the R\'{e}nyi entropy response is only related to the spectral functions of the system and the environment, as well as a specific statistical kernel function. General properties of bosonic systems and fermionic systems are discussed and we predict a short time quadratic growth (decay) and long time linear growth (decay). 

Further, we apply the linear response theory to Sachdev-Ye-Kitaev (SYK) model\cite{Kitaev15, SY, Maldacena16, Kitaev17} with a free fermion bath with arbitrary different temperatures. SYK model is recently discovered as an example of holographic duality\cite{Kitaev17, Maldacena16a, Gross17}. It is dual to a Jackiw-Teitelboim (JT) gravity\cite{Jackiw,Teitelboim} in AdS$_2$ spacetime\cite{Kitaev17,Maldacena16}. Since SYK model is a solvable model in large N limit, quite a few results for SYK model's entropy dynamics are obtained for various SYK-like models\cite{Gu17a,Gu17b,Balents17,Pengfei19,Yiming20,Pengfei20a,Pengfei20b,Banerjee20,Pengfei20c} and for Maldacena-Qi model\cite{MQ} which is dual to a traversable wormhole. These entropy dynamics are also calculated in gravity side\cite{Penington19, Almheiri19, Maldacena20, Pedraza17,Yiming19}. Here we try to ask, to what kind of extent could our perturbative theory apply to a general case. Can we recover the Page curve in `` black hole '' evaporation case\cite{Page01,Page02} where SYK model's initial temperature is higher than the environment's temperature. A Page curve is an  entanglement entropy curve of the environment which follows a linear decay after an initial linear growth. Even through we are calculating black hole's entropy, still, we wonder if we can see signals for Page curve.  Eventually, we observed this Page curve like behavior in long time dynamics. Further we can change a gapless fermion to a gapped fermion to change the environment's spectrum dramatically. an interesting phenomenon is observed for RE oscillation at some special temperature window. It seems that the entangled information can travel back and forth between the system and the bath in this situation, which is quite curious. 

Our paper is orgarnized as follows. In the next section, we first establish the general linear response theory for RE of the subsystem. Then we discuss general properties of RE dynamics for bosonic systems and fermionic systems. Then in the third section, we apply our theory to SYK model with a non-Markovian bath and we changed the temperature and the spectrum of environment to calculate the RE dynamics. Finally, we summarize our results in the conclusion section.

%To summarize, in this new theory, previous LRT theories are unified and some new response theories are proposed. Hopefully, we can make use of these new theories to measure information structure of quantum matters, which is quite difficult in previous experimental protocols. The whole paper are divided into two parts. In the first part, we present a general theory for the response theory of observables, retard correlation function, and Renyi entropy. In the second part, we apply our theory for observables response to free fermions system, theory for correlation function to Maldacena-Qi model and theory for Renyi entropy response to coupled Sachdev-Ye-Kitaev (SYK) models which can be interpreted to black hole evaporation. These examples are presented in section 3 to section 5.  Fianlly, we discuss the connection between our theory and the condensed matter, cold atoms experiments.

\section{General Entropy Response Theory}
Here we consider a general situation for a sudden coupling between a system and a bath. Before the coupling, the hamiltonian is $\hat{H}_0=\hat{H}_{\rm S}+\hat{H}_{\rm E}$, and after the coupling the hamiltonian becomes
\ba
\hat{H}=\hat{H}_{\rm S}+\hat{H}_{\rm E}+\hat{V}.
\ea
Here $\hat{V}=g \sum_j(\hat{O}_j^\dag\hat{\xi}^{}_j+\hat{\xi}_j^\dag\hat{O}_j^{})$, where $\hat{O}_j$ is an operator acting on the system and $\hat{\xi}_j$ is an operator in the environment. $j$ is the index of the operator. We can find out that the system's density operator evolutes like follows
\begin{eqnarray}
\hat{\rho}_{\rm S}(t)={\rm Tr}_E\left(e^{-i\hat{H}t}\hat{\rho}_0^{\rm tot} e^{i\hat{H}t}\right)/{\rm Tr}(\hat{\rho}_{\rm E}(t)),\\
\hat{\rho}_{\rm E}(t)={\rm Tr}_S\left(e^{-i\hat{H}t}\hat{\rho}_0^{\rm tot} e^{i\hat{H}t}\right)/{\rm Tr}(\hat{\rho}_{\rm S}(t)),
\end{eqnarray}
where $\hat{\rho}_0^{\rm tot}$ is the initial density operator, which we assume as a product state of thermal states in the system and in the bath. 
\ba
\hat{\rho}_0^{\rm tot}=\hat{\rho}_0\otimes \hat{\rho}_{E}.
\ea
Because after a unitary evolution, the trace of $\hat{\rho}_{\rm S}(t)$ and $\hat{\rho}_{\rm E}(t)$ is invariant. Therefore we can simply replace ${\rm Tr}_E(\hat{\rho}_{\rm E}(t))$ to $Z_{\rm E}={\rm Tr}_E(\hat{\rho}_{\rm E}(0))$.  Here we employ ${\rm Tr}_E$ as a trace over the environment's Hilbert space. ${\rm Tr}$ is short for ${\rm Tr}_S$ as trace over system's Hilbert space. 

The reduced density operator of the system is evoluting in an interacting picture as 
\ba
\hat{\rho}_{\rm S}(t)&=&{\rm Tr}_E\left(e^{i\hat{H}t}e^{-i\hat{H}_0 t}e^{i\hat{H_0}t}\hat{\rho}_0\otimes\hat{\rho}_{\rm E}e^{-i\hat{H}_0 t}e^{i\hat{H}_0 t}e^{-i\hat{H}t}\right)/Z_{\rm E}\nonumber\\
&\equiv&{\rm Tr}_E\left({\cal U}(0,t)\hat{\rho}^{\rm I}_0(t)\otimes\hat{\rho}_{\rm E}^{\rm I}(t){\cal U}(t,0)\right)/Z_{\rm E}\label{rhos}
\ea
Here we introduce the evolution operator ${\cal U}(0,t)\equiv e^{i\hat{H}_0 t}e^{-i\hat{H}t}=\hat{\cal T}_t\exp(-i\int_0^t \hat{V}^{\rm I}(t')dt')$, where $\hat{V}^{\rm I}(t')=e^{i\hat{H}_0 t}\hat{V}e^{-i\hat{H}_0t}$ is in interaction picture, $\hat{\cal T}_t$ is time-ordering operator.
%\ba
%\hat{\rho}_{\rm S}(t)&=&{\rm Tr}_E\left((1-i\int_0^t \hat{V}(t')dt'-\int_0^t dt_1\int_0^{t_1} dt_2 \hat{V}(t_1)\hat{V}(t_2))\hat{\rho}_0\right.\nonumber\\
%&&\left.(1+i\int_0^t \hat{V}(t')dt'-\int_0^t dt_1\int_0^{t_1} dt_2 \hat{V}(t_2)\hat{V}(t_1))\right)/Z_{\rm E}\nonumber\\
%&=&\hat{\rho}_0+g^2\int_0^t dt_1\int_0^{t_1}dt_2\langle \hat{\xi}_j(t_1)\hat{\xi}_\ell^\dag(t_2)\rangle_{\rm E}\left(\hat{\cal O}^\dag_j(t_1)\hat{\rho}_0\hat{\cal O}_\ell(t_2)-\hat{\cal O}^\dag_j(t_1)\hat{\cal O}^{}_{\ell}(t_2)\hat{\rho}_0\right)\nonumber\\
%&&\hspace{2.95ex}+g^2\int_0^t dt_1\int_0^{t_1}dt_2\langle \hat{\xi}_j^\dag(t_1)\hat{\xi}_\ell^{}(t_2)\rangle_{\rm E}\left(\hat{\cal O}^{}_j(t_1)\hat{\rho}_0\hat{\cal O}_\ell^\dag(t_2)-\hat{\cal O}^{}_j(t_1)\hat{\cal O}^{\dag}_{\ell}(t_2)\hat{\rho}_0\right)\nonumber\\
%&&\hspace{2.95ex}+g^2\int_0^t dt_1\int_0^{t_1}dt_2\langle \hat{\xi}_j(t_2)\hat{\xi}_\ell^\dag(t_1)\rangle_{\rm E}\left(\hat{\cal O}^\dag_j(t_2)\hat{\rho}_0\hat{\cal O}_\ell(t_1)-\hat{\rho}_0\hat{\cal O}^\dag_j(t_2)\hat{\cal O}^{}_{\ell}(t_1)\right)\nonumber\\
%&&\hspace{2.95ex}+g^2\int_0^t dt_1\int_0^{t_1}dt_2\langle \hat{\xi}_j^\dag(t_2)\hat{\xi}_\ell^{}(t_1)\rangle_{\rm E}\left(\hat{\cal O}^{}_j(t_2)\hat{\rho}_0\hat{\cal O}_\ell^\dag(t_1)-\hat{\rho}_0\hat{\cal O}^{}_j(t_2)\hat{\cal O}^{\dag}_{\ell}(t_1)\right)
%\ea
By inserting $\hat{\cal U}(0,t)$ and $\hat{\cal U}(t,0)=\hat{\cal U}(0,t)^\dag$ into Eq.~(\ref{rhos}), keeping to $g^2$ order, we have
\ba
\hat{\rho}_{\rm S}(t)&=&{\rm Tr}_E\left((1-i\int_0^t \hat{V}^{\rm I}(t')dt'-\int_0^t dt_1\int_0^{t_1} dt_2 \hat{V}^{\rm I}(t_1)\hat{V}^{\rm I}(t_2))\hat{\rho}_0\otimes\hat{\rho}_{\rm E}\right.\nonumber\\
&&\left.(1+i\int_0^t \hat{V}^{\rm I}(t')dt'-\int_0^t dt_1\int_0^{t_1} dt_2 \hat{V}^{\rm I}(t_2)\hat{V}^{\rm I}(t_1))\right)/Z_{\rm E}.
\ea
More explicitly,
\ba
\hat{\rho}_{\rm S}(t)&=&\hat{\rho}_0+g^2\int_0^t dt_1\int_0^{t_1}dt_2\langle \hat{\xi}^{\rm I}_j(t_1)\hat{\xi}_\ell^{{\rm I},\dag}(t_2)\rangle_{\rm E}\left(\hat{\cal O}^{\rm I}_\ell(t_2)\hat{\rho}_0\hat{\cal O}_j^{{\rm I},\dag}(t_1)-\hat{\cal O}^{{\rm I},\dag}_j(t_1)\hat{\cal O}^{\rm I}_{\ell}(t_2)\hat{\rho}_0\right)\nonumber\\
&&\hspace{2.95ex}+g^2\int_0^t dt_1\int_0^{t_1}dt_2\langle \hat{\xi}_\ell^{{\rm I},\dag}(t_2)\hat{\xi}_j^{\rm I}(t_1)\rangle_{\rm E}\left(\hat{\cal O}^{{\rm I},\dag}_j(t_1)\hat{\rho}_0\hat{\cal O}_\ell^{\rm I}(t_2)-\hat{\cal O}^{\rm I}_\ell(t_2)\hat{\cal O}^{{\rm I},\dag}_{j}(t_1)\hat{\rho}_0\right)\nonumber\\
&&\hspace{2.95ex}+g^2\int_0^t dt_1\int_0^{t_1}dt_2\langle \hat{\xi}^{\rm I}_j(t_2)\hat{\xi}_\ell^{{\rm I},\dag}(t_1)\rangle_{\rm E}\left(\hat{\cal O}^{\rm I}_\ell(t_1)\hat{\rho}_0\hat{\cal O}_j^{{\rm I},\dag}(t_2)-\hat{\rho}_0\hat{\cal O}^{{\rm I},\dag}_j(t_2)\hat{\cal O}^{\rm I}_{\ell}(t_1)\right)\nonumber\\
&&\hspace{2.95ex}+g^2\int_0^t dt_1\int_0^{t_1}dt_2\langle \hat{\xi}_j^{{\rm I},\dag}(t_2)\hat{\xi}_\ell^{\rm I}(t_1)\rangle_{\rm E}\left(\hat{\cal O}^{{\rm I},\dag}_\ell(t_1)\hat{\rho}_0\hat{\cal O}_j^{\rm I}(t_2)-\hat{\rho}_0\hat{\cal O}^{\rm I}_j(t_2)\hat{\cal O}^{{\rm I},\dag}_{\ell}(t_1)\right),\nonumber
\ea
where $\langle\cdot\rangle_{\rm E}={\rm Tr}_E(\hat{\rho}_{\rm E}\cdot)/Z_{\rm E}$ is the environment ensemble average. Operators with upper index ${\rm I}$ are in interacting picture.
By introducing Green's function for $\hat{\xi}$ field and Lindbald-like operators $\hat{\cal L}^{d,e}_{j\ell,{\cal O}}$, we can simplify the above result as
\ba
\hat{\rho}_{\rm S}(t)=\hat{\rho}_0\!+\!ig^2\!\!\iint_0^t \!\!\! dt_1dt_2  \left(G^<_{\xi, j\ell}(t_1,t_2)\hat{\cal L}^d_{j\ell,{\cal O}}(t_1,t_2)\!+\!G^>_{\xi, j\ell}(t_1,t_2)\hat{\cal L}^{e}_{j\ell,{\cal O}}(t_1,t_2)\right)\!\hat{\rho}_0,\label{rhosA}
\ea
\begin{figure}[h]\centering
\includegraphics[width=15cm]{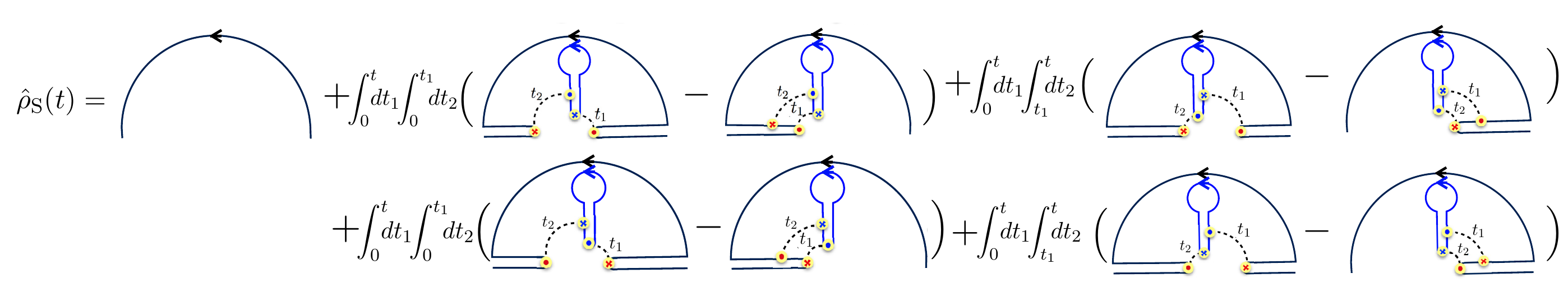}
\caption{Diagrammatic illustration of Eq.~(\ref{rhosA})}\label{FigRhos}
\end{figure}
where
\ba
G^<_{\xi,j\ell}(t_1,t_2)\equiv -i\langle \hat{\xi}_j^{\rm I}(t_1)\hat{\xi}_\ell^{{\rm I},\dag}(t_2)\rangle_{\rm E},\hspace{3ex}G^>_{\xi,j\ell}(t_1,t_2)\equiv -i\langle \hat{\xi}^{{\rm I},\dag}_\ell(t_1)\hat{\xi}_j^{\rm I}(t_2)\rangle_{\rm E},
\ea
$\hat{\cal L}^d$ and $\hat{\cal L}^e$ are dissipation and enhancement operators which are defined as 
\ba
\hat{\cal L}^d_{j\ell,{\cal O}}(t_1,t_2)\hat{\rho}_0&\equiv&\hat{\cal O}^{\rm I}_\ell(t_2) \hat{\rho}_0\hat{\cal O}^{{\rm I},\dag}_j(t_1)\!-\!\theta_{12}\hat{\cal O}^{{\rm I},\dag}_j(t_1)\hat{\cal O}^{\rm I}_{\ell}(t_2)\hat{\rho}_0\!-\!\theta_{21}\hat{\rho}_0\hat{\cal O}^{{\rm I},\dag}_j(t_1)\hat{\cal O}^{\rm I}_{\ell}(t_2),\\
\hat{\cal L}^e_{j\ell,{\cal O}}(t_1,t_2)\hat{\rho}_0&\equiv& \hat{\cal O}_\ell^{{\rm I},\dag}(t_2)\hat{\rho}_0 \hat{\cal O}^{\rm I}_j(t_1)\!-\!\theta_{12}\hat{\cal O}^{\rm I}_j(t_1)\hat{\cal O}^{{\rm I},\dag}_{\ell}(t_2)\hat{\rho}_0\!-\!\theta_{21}\hat{\rho}_0\hat{\cal O}^{\rm I}_j(t_1)\hat{\cal O}^{{\rm I},\dag}_{\ell}(t_2).
\ea
Here $\theta_{12}=\theta(t_1-t_2)$ is heavyside step function, which is 1 when $t_1>t_2$, $-1$ when $t_1<t_2$, and it is $\frac{1}{2}$ when $t_1=t_2$. Here we also assumed that $\langle\hat{\xi}(t)\rangle_{\rm E}=0$. This condition is usually satisfied. Here we assume the vacuum of environment is not a symmetry broken state of $\hat{\xi}_j$. 

Now we introduce a diagram to describe above result. Diagrammatically, Eq.~(\ref{rhosA}) can be expressed as Fig.~\ref{FigRhos}. The diagram rules are the following, we use red dot to represent $\hat{\cal O}^\dag$, red cross to represent $\hat{\cal O}^{}$. Here we dropped the subindices of operators $\hat{\cal O}$. Blue dot is $\hat{\xi}^\dag$ and blue cross is $\hat{\xi}$. The arrow lines are evolution operators. The half circle represent an evolution in imaginary time in the system --- $e^{-\beta\hat{H}_{\rm S}}$ and the blue full circle is $e^{-\beta_{\rm E}\hat{H}_{\rm E}}$. A closed circle means trace. Every dashed line connecting blue dot(cross) and red cross(dot) carries a factor $g$. The length of lines in x direction represents real time, representing $e^{-i\hat{H}t}$ when the arrow is pointing at the origin of the circle and representing $e^{i\hat{H}t}$ in outgoing direction. A closed blue curve represents $G^{>,<}_\xi$. One can find, by applying these rules, the diagrams given in Fig.~\ref{FigRhos} recover Eq.~(\ref{rhosA}).

\begin{figure}[h]\centering
\includegraphics[width=13cm]{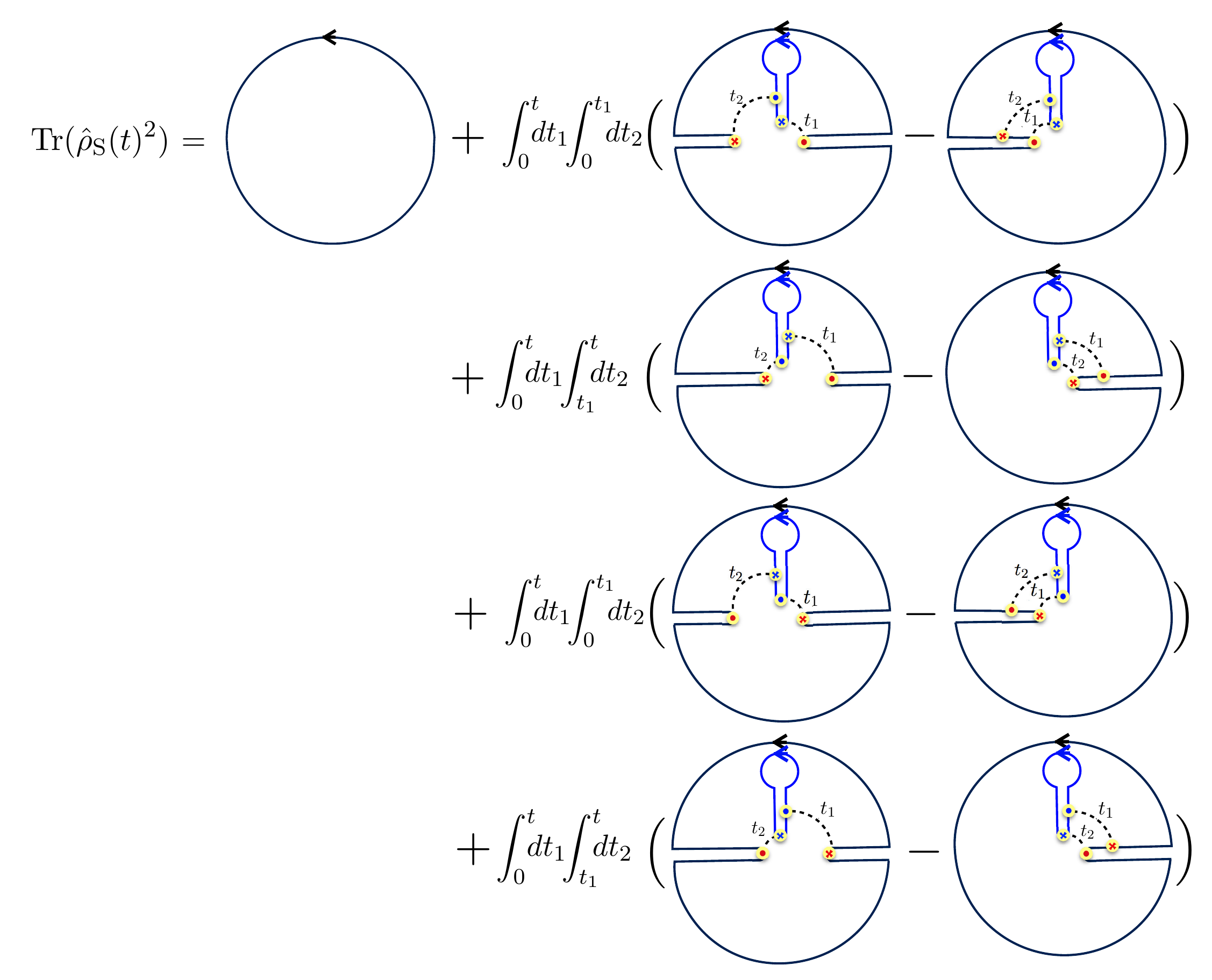}
\caption{Perturbation digrams for ${\rm Tr}(\hat{\rho}_{\rm S}^2(t))$, up to $g^2$ order. Diagram representation of Eq.~(\ref{RE1})}\label{RhosSq}
\end{figure}

Then we are ready to calculate the R\'{e}nyi entropy of the system up to $g^2$ order. Now we calculate the second R\'{e}nyi entropy, which is $S_{\rm RE}^{(2)}=-\log({\rm Tr}(\hat{\rho}_{\rm S}^2(t)))$. Here ${\rm Tr}$ is trace over system. First we calculate ${\rm Tr}(\hat{\rho}_{\rm S}(t)^2)$,
\ba
{\rm Tr}(\hat{\rho}_{\rm S}(t)^2)&=&{\rm Tr}(\hat{\rho}_0^2)+2ig^2\sum_{j,\ell}\iint_0^t \!\!\! dt_1dt_2 \left(G^<_{\xi, j\ell}(t_1,t_2){\rm Tr}(\hat{\rho}_0(\hat{\cal L}^d_{j\ell,{\cal O}}(t_1,t_2)\hat{\rho}_0))\right.\nonumber\\
&&\left.\hspace{27.ex}+G^>_{\xi, j\ell}(t_1,t_2){\rm Tr}(\hat{\rho}_0(\hat{\cal L}^{e}_{j\ell,{\cal O}}(t_1,t_2)\hat{\rho}_0))\right).\label{RE1}
\ea
Above expression can also displayed in diagram language as is shown in Fig.~\ref{RhosSq}.

Now we discuss two situations, the first case is $\hat{\cal O}_j$ and $\hat{\xi}_j$ are both bosonic annihilation operators. Both in Green's function and the $\hat{\cal L}$ operators, there are sub-indices $j$ and $\ell$.  Here we assume $G^{>,<}_{\xi, j\ell}=G^{>,<}_{\xi,j}\delta_{j\ell}$. Then only $\hat{\cal L}^{d,e}_{j\ell,{\cal O}}$ contributes. In this diagonal basis, we assume $G^{>,<}_{\xi,j}$ and $\hat{\cal L}^{d,e}_{j,{\cal O}}$ are $j$ independent, then we denote $G^{>,<}_{\xi,j}$ as$G^{>,<}_{\xi}$ and $\hat{\cal L}^{d,e}_{j,{\cal O}}$ as $\hat{\cal L}^{d,e}_{{\cal O}}$ for simplicity. According to the relation between the greater (lesser) Green's function and the spectral function, we have
\ba
G_\xi^>(t_1,t_2)=-i\int d\omega'{\cal A}_\xi(\omega')\frac{1}{e^{\beta_E \omega'}-1}e^{i\omega'(t_1-t_2)},\label{greater}
\ea
\ba
G_\xi^<(t_2,t_1)=-i\int d\omega'{\cal A}_\xi(\omega')\frac{e^{\beta_E \omega'}}{e^{\beta_E \omega'}-1}e^{i\omega'(t_1-t_2)},\label{lesser}
\ea
where ${\cal A}_\xi(\omega)$ is the spectral function of $\hat{\xi}$ field, ${\cal A}_\xi(\omega)=-\frac{1}{\pi}{\rm Im}G^R_\xi(\omega)$ where $G_\xi^R(t)=-i\theta(t)\langle [\hat{\xi}(t),\xi^\dag(0)]\rangle$ is the retard Green's function of $\hat{\xi}$ field. $1/\beta_E$ is the environment temperature. Here we can see, a white noise bath can be generated in two cases. At high temperature, $\beta_E$ is close to zero, therefore when ${\cal A}_\xi(\omega)\sim\omega$, $G_\xi^{>,<}\propto \delta(t_1-t_2)$. This is Ohm bath at high temperature. At low temperature, we find $G^>_\xi\sim 0$, and a white noise is generated by a constant density of states ${\cal A}_\xi\sim {\rm const.}$. This is the other case of white noise. 

On the other hand, we find ${\rm Tr}(\hat{\rho}_0\hat{\cal L}^d_{{\cal O}}(t_1,t_2))={\rm Tr}(\hat{\rho}_0\hat{\cal L}^e_{{\cal O}}(t_2,t_1))$ is a two-point Wightman Green's function,
\ba
{\rm Tr}\left(\hat{\rho}_0\hat{\cal O}^\dag(t_1)\hat{\rho}_0\hat{\cal O}(t_2)\right)={\rm Tr}\left(\hat{\rho}_0 \hat{\cal O}^\dag \hat{\rho}_0 \hat{\cal O}(t_2-t_1)\right)=\int\frac{d\omega}{2\pi}{\cal A}_{\cal O}(\omega)\frac{e^{\beta\omega}}{e^{2\beta \omega}-1}e^{i\omega(t_2-t_1)},\label{Wightman}
\ea
where ${\cal A}_{\cal O}(\omega)$ is the spectral function of ${\cal O}$ field. The last equation in above line can be proved by Lehmann spectrum Theorem. By inserting Eq.~(\ref{greater}), Eq.~(\ref{lesser}) and Eq.~(\ref{Wightman}) into Eq.~(\ref{RE1}), the deviation of Renyi entropy $\delta S_{\rm RE}^{(2)}(t)\equiv S_{\rm RE}^{(2)}(t)+\log({\rm Tr}(\hat{\rho}_0^2))$ is
\ba
\delta S_{\rm RE}^{(2)}(t)=\frac{2g^2 N_j}{{\rm Tr}(\rho_0^2)}\int_0^t dt_0 d\omega d\omega' \frac{\sin(\omega-\omega')t_0}{\omega-\omega'}{\cal W}_{\beta\beta_E}^B(\omega,\omega'){\cal A}_\xi(\omega'){\cal A}_{\cal O}(\omega)\label{REBOSE}
\ea
where $N_j$ is the mode number, and 
\ba
{\cal W}_{\beta\beta_E}^B(\omega,\omega')=\frac{(e^{\beta\omega}-1)(e^{\beta\omega}-e^{\beta_E\omega'})}{(e^{2\beta\omega}-1)(e^{\beta_E\omega'}-1)}.
\ea
At high temperature limit and ${\cal A}_\xi$ being Ohm bath, we have $\delta S_{\rm RE}^{(2)}\sim \lambda_0 t$ where $\lambda_0=2g^2/{\rm Tr}(\rho_0^2)\int d\omega (e^{\beta\omega}-1){\cal A}_{\cal O}(\omega)/(e^{\beta\omega}+1)$ is the entropy growth rate. According to this formula, we learn, the entropy response is only dependent on the distribution function and the spectrum of bath and the system.   

As the entropy response theory at $g^2N_j$ order is indeed an exponential contribution in ${\rm Tr}(\hat{\rho}^2_S(t))$'s perturbation, therefore the response theory for entropy may have larger applicable region. Actually, we can apply a perturbation calculation to next order $g^4$, where we can capture how the spectrum function of the system and the environment is changed by interactions. But these back action effects are not considered in present calculation.

 Quite similarly, we can work out the R\'{e}nyi entropy response for fermionic bath and fermionic operator $\hat{\cal O}$, which is
 \ba
\delta S_{\rm RE}^{(2)}(t)=\frac{2g^2 N_j}{{\rm Tr}(\rho_0^2)}\int_0^t dt_0 d\omega d\omega' \frac{\sin(\omega-\omega')t_0}{\omega-\omega'}{\cal W}_{\beta\beta_E}^F(\omega,\omega'){\cal A}_\xi(\omega'){\cal A}_{\cal O}(\omega),\label{REFERMI}
\ea
where
\ba
{\cal W}_{\beta\beta_E}^F(\omega,\omega')=\frac{(e^{\beta\omega}-1)(e^{\beta\omega}-e^{\beta_E\omega'})}{(1+e^{2\beta\omega})(1+e^{\beta_E\omega'})}.
\ea
Notice that in fermionic case, the Markovian condition is requiring a frequency independent spectrum together with frequency independent distribution function. The latter condition is satisfied only for zero and infinite temperature. Therefore there are two ways to break down Markovian approximation, one is by finite temperature; another is by adding strong frequency dependence in the noise spectrum of the environment.

\subsection{General Features for Bosonic systems}

In this section, we first discuss the short time and long time behavior of entanglement entropy growth for bosonic environment. Then we derive a `` Fourier law '' in heat transport when the system and the environment have small temperature difference.

First, we discuss the short time and long time dynamics of $\delta S_{\rm RE}^{(2)}$. When $t$ is small, in a sense ${\cal J}t\ll 1$, ${\cal J}'t\ll1$ ( ${\cal J}$ and ${\cal J}'$ are typical energy scale for the system and the bath respectively ), we have
\ba
\frac{\sin(\omega-\omega')t_0}{\omega-\omega'}\approx t_0\nonumber
\ea
For this reason, the short time dynamics is always proportional to $t^2$. 
\ba
\delta S_{\rm RE}^{(2)}(t)=\frac{1}{2}\kappa_{\rm B}t^2
\ea
where coefficient is 
\ba
\kappa_{\rm B}\equiv\frac{2g^2 N_j}{{\rm Tr}(\rho_0^2)}\int d\omega\int d\omega' {\cal W}_{\beta\beta_E}^B(\omega,\omega'){\cal A}_{\cal O}(\omega){\cal A}_{\xi}(\omega').\label{kappaB}
\ea
Bosonic environment is unique because the chemical potential of Bose distribution is negative ( although in our formula we take chemical potential to be zero ), and the frequency of bosonic excitations are always positive. Therefore when $\beta_{\rm E}>\beta$, and the environment has a lower temperature, $\kappa_{\rm B}$ is usually negative. That is due to ${\cal A}_{\cal O}(\omega)$, ${\cal A}_\xi(\omega)$ are positive spectral functions. Further
\ba
{\cal W}_{\beta\beta_E}^B(\omega,\omega')=\frac{e^{\beta\omega}-e^{\beta_{\rm E}\omega'}}{(e^{\beta\omega}+1)(e^{\beta_{\rm E}\omega'}-1)}\approx\frac{e^{\beta\omega-\beta_{\rm E}\omega'}-1}{e^{\beta\omega}+1}\nonumber
\ea
As $\omega$, $\omega'$ are positive, as long as $\beta\omega<\beta_{\rm E}\omega'$, the factor
${\cal W}_{\beta\beta_E}^B(\omega,\omega')$ is negative. As long as the low energy spectrum of the bath and the system are large, then $\kappa_{\rm B}<0$. Later we will see it is completely different in fermionic bath case.

Now we turn to long time case. Here we assume time is long and ${\cal J}t_0\sim {\cal J}'t_0\sim 1$, we have 
\ba
\frac{\sin(\omega-\omega')t_0}{\omega-\omega'}\sim\delta(\omega-\omega')\nonumber
\ea
Therefore the long time dynamics is approximately
\ba
\delta S_{\rm RE}^{(2)}(t)=\frac{2g^2 N_j}{{\rm Tr}(\rho_0^2)}t \int d\omega {\cal W}_{\beta\beta_E}^B(\omega,\omega){\cal A}_\xi(\omega){\cal A}_{\cal O}(\omega)\label{REBOSE_A}
\ea
The sign for the entropy growth rate in long time limit is determined by the sign of ${\cal W}_{\beta\beta_E}^B(\omega,\omega)$.  ${\cal W}_{\beta\beta_E}^B(\omega,\omega)>0$ when $\beta_{\rm E}<\beta$. In this situation the environment's temperature is higher than the system. ${\cal W}_{\beta\beta_E}^B(\omega,\omega)<0$ when $\beta_{\rm E}>\beta$. In this situation, the environment has a lower temperature than the system. 

Now we are going to derive a Fourier heat transport formula from our Eq.~(\ref{REBOSE_A}). Now we assume two temperature $T=1/\beta$ and $T_{\rm E}=1/\beta_{\rm E}$ are close. Here we have taken $k_{\rm B}=1$. Further we assume $\beta_{\rm E}$ is small, then we have
\ba
{\cal W}_{\beta\beta_E}^B(\omega,\omega)=\frac{1}{e^{\beta\omega}+1}\frac{e^{(\beta-\beta_{\rm E})\omega}-1}{1-e^{-\beta_{\rm E}\omega}}\approx\frac{T_{\rm E}-T}{T}\frac{1}{(e^{\beta\omega}+1)}\label{BoseW}
\ea
Insert Eq.~(\ref{BoseW}) into Eq.~(\ref{REBOSE_A}), we have
\ba
T\delta S_{\rm RE}^{(2)}(t)=\frac{2g^2 N_j}{{\rm Tr}(\rho_0^2)}\left(\int_0^\infty d\omega\frac{{\cal A}_{\cal O}(\omega){\cal A}_\xi(\omega)}{e^{\beta\omega}+1}\right)(T_{\rm E}-T)t
\ea
If we can define $dQ\equiv T\delta S_{\rm RE}^{(2)}(t)$ as some kind of heat, therefore above equation shows a Fourier law of heat transport. The heat transport rate is proportional to the temperature difference between the system and the bath. If we define $F_{\rm Q}=dQ/dt/(T_{\rm E}-T)$, then the transport coefficient $F_{\rm Q}$ is predicted as
\ba
F_{Q}=\frac{2g^2 N_j}{{\rm Tr}(\rho_0^2)}\int_0^\infty d\omega\frac{{\cal A}_{\cal O}(\omega){\cal A}_\xi(\omega)}{e^{\beta\omega}+1}.
\ea

\subsection{General Features for Fermionic systems}

In this section, we discuss the short time and long time entropy growth for fermionic bath. Meanwhile we will give the `` heat transport '' coefficient for fermionic bath.

First we discuss the short time behavior of $\delta S_{\rm RE}^{(2)}(t)$. Similar to the case of bosonic system, we have
\ba
\delta S_{\rm RE}^{(2)}(t)=\frac{1}{2}\kappa_{\rm F}t^2,
\ea
where
\ba
\kappa_{\rm F}=\frac{2g^2 N_j}{{\rm Tr}(\rho_0^2)}\int d\omega\int d\omega' {\cal W}_{\beta\beta_E}^F(\omega,\omega'){\cal A}_{\cal O}(\omega){\cal A}_{\xi}(\omega').
\ea
The only difference with Eq.~(\ref{kappaB}) is the distribution function ${\cal W}_{\beta\beta_E}^B$ is replaced by ${\cal W}_{\beta\beta_E}^F$. For fermions, negative frequency is meaningful because of the presence of Fermi level. Here Fermi level is taken to be zero.  At low temperature,
\ba
{\cal W}_{\beta\beta_E}^F(\omega,\omega')\approx\frac{(e^{\beta\omega}-1)(e^{\beta\omega-\beta_{\rm E}\omega'}-1)}{e^{2\beta\omega}+1}
\ea
It is clear that when $\omega<0$, $\omega'>0$, ${\cal W}_{\beta\beta_E}^F(\omega,\omega')>0$. Therefore $\kappa_{\rm F}$ is usually positive rather than being negative. This is very different from boson case. 

Now we consider the long time behavior of $\delta S_{\rm RE}^{(2)}(t)$. In long time limit, we have
\ba
\delta S_{\rm RE}^{(2)}(t)=\frac{2g^2 N_j}{{\rm Tr}(\rho_0^2)}t \int d\omega {\cal W}_{\beta\beta_E}^F(\omega,\omega){\cal A}_\xi(\omega){\cal A}_{\cal O}(\omega).\label{REFERMI_A}
\ea
One can prove that for $\beta_{\rm E}>\beta$, ${\cal W}_{\beta\beta_E}^F(\omega,\omega)<0$ for any $\omega$. Therefore if the environment has a larger initial temperature, $\delta S_{\rm RE}^{(2)}(t)$ experienced a linear growth and if the environment temperature is lower than the system, $\delta S_{\rm RE}^{(2)}(t)$ is decreasing linearly in long time evolution. However, as we state previously, the short time dynamics of low temperature fermionic bath is an entropy growth, therefore the entropy dynamics could be non-monotonic. Later in the next section, we will show this case in a specific example. Here we stress that the kink in $\delta S_{\rm RE}^{(2)}(t)$ could be very general in fermionic case, not necessarily as an interpretation of gravity duality. 

Finally, we discuss the situation when the system and the bath's temperature difference is small. Again a Fourier-alike law can be presented,
\ba
{\cal W}_{\beta\beta_E}^F(\omega,\omega)=\frac{e^{\beta\omega}-1}{e^{2\beta\omega}+1}\frac{e^{(\beta-\beta_{\rm E})\omega}-1}{1+e^{-\beta_{\rm E}\omega}}\approx\frac{(e^{\beta\omega}-1)\omega\beta\beta_{\rm E}}{(e^{2\beta\omega}+1)(1+e^{-\beta_{\rm E}\omega})}(T_{\rm E}-T)\nonumber
\ea
Again we can get a `` Fourier law '' in `` heat transport '' if we define $T\delta S_{\rm RE}^{(2)}(t)$ as `` heat flow '' --- $dQ$. Again we define $F_{\rm Q}=dQ/dt/(T_{\rm E}-T)$, then
\ba
F_{\rm Q}=\frac{2g^2 N_j}{{\rm Tr}(\rho_0^2)}\int_0^\infty d\omega\frac{{\cal A}_{\cal O}(\omega){\cal A}_\xi(\omega)(e^{\beta\omega}-1)\omega\beta_{\rm E}}{(e^{2\beta\omega}+1)(1+e^{-\beta_{\rm E}\omega})}.
\ea
One can observe that the temperature dependence in $F_{\rm Q}$ is very different from bosonic bath case.

\section{Renyi Entropy Response in SYK model}

In this section, we apply the general R\'{e}nyi entropy response theory to Sachdev-Ye-Kitaev (SYK) model with an non-Markovian environment. We assume the system is SYK$_4$ whose hamiltonian is,
\ba
\hat{H}_{\rm S}=\sum_{jklm}J_{jklm}\hat{\chi}_j\hat{\chi}_k\hat{\chi}_l\hat{\chi}_m,
\ea
where $\hat{\chi}_j$ is Majorana fermion operator. $J_{jklm}$ are random, and we have
\ba
\overline{J_{jklm}}=0,\hspace{4ex}\overline{J_{jklm}^2}=\frac{3!}{N^3}{\cal J}^2,
\ea
where $\overline{\cdot}$ is disorder average, $N$ is the total modes number. Throughout our discussion we are in conformal limit of SYK model, so that $N\gg\beta{\cal J}\gg 1$.

Here we are going to consider two kinds of baths. One bath is a SYK$_2$ model with zero chemical potential. In this model, when the coupling strength is large, then the spectral function around zero frequency is a constant. This kind of bath has a Markovian limit. Another bath is two SYK$_2$ model with opposite on site energies, which is more like a gapped system and is far away from Markovian environment. Here we stress that non-Markovian effect comes from both the frequency dependence in spectral function of the environment as well as the the distribution function kernel ${\cal W}^{F}$. Therefore we separate these two factors with a constant spectral function and a strongly frequency dependent spectral fucntion. In the former, the non-Markovian effect comes from the distribution function, while in the latter, the non-Markovian effect comes from the spectral function.

\subsection{Heating and Evaporation with Markovian-like Environment}
Here we assume the environment is random free fermions, SYK$_2$, whose hamiltonian is
\ba
\hat{H}_{\rm E}=\sum_{\alpha=1}^M\sum_{jk=1}^{N} J'_{jk}\hat{\psi}_{j,\alpha}^\dag\hat{\psi}_{k,\alpha}^{},
\ea
where $\hat{\psi}_j$ are complex fermion annihilation operators. $J'_{jk}$ is random, and it satisfies
\ba
\overline{J'_{jk}}=0,\hspace{4ex}\overline{(J_{jk}^{\prime})^2}=\frac{1}{N}({\cal J}^{\prime})^2
\ea
The interaction between the system and the bath is $\hat{V}$,
\ba
\hat{V}=g\sum_{j=1}^N\sum_{\alpha=1}^M (\hat{\chi}_j\hat{\psi}^\dag_{j,\alpha}+\hat{\psi}^{}_{j,\alpha}\hat{\chi}_j)
\ea
One advantage is that SYK model is known to have a gravitaional duality. SYK model's low energy effective action can be described by a Schwarzian theory, which is dual to the Jackiw-Teitelboim (JT) gravity theory in AdS$_2$ spacetime\cite{Kitaev17,Maldacena16}. The present situation of sudden coupling between the SYK$_4$ and SYK$_2$ model can be viewed as an black hole evaporation problem if the environment's temperature is much lower than the black hole. In gravitational side, it is known that evaporation experience an entanglement entropy growth and drop, which is well-known as Page curve. The entanglement entropy calculation based on RT formula is done recently\cite{Penington19,Almheiri19,Maldacena20} towards an understanding of Page curve.  Therefore, according to holographic principle, we expect similar non-monotonic behavior in SYK$_4$'s entanglement entropy dynamics.

Now we apply the perturbation theory to calculate the R\'{e}nyi entropy dynamics after a sudden coupling between SYK$_4$ and SYK$_2$ by $\hat{V}$ to see how the second R\'{e}nyi entropy response to the sudden coupling between these two systems.

From our formula, we find the most important information for R\'{e}nyi entropy response is the spectrum of the system and the spectrum of the environment. These spectrum can be obtained by calculating retard Green's function. Compared with the general theory, we find $\hat{\cal O}_j=\hat{\chi}_j$ and $\hat{\xi}_j=\hat{\psi}_{j,\alpha}$. Then the spectral function for $\hat{\cal O}_j$ field is
\ba
{\cal A}_{\cal O}(\omega)=\frac{b}{\sqrt{\beta}}\frac{\Gamma(\frac{1}{4}-i\beta\omega/2\pi)}{\Gamma(\frac{3}{4}-i\beta\omega/2\pi)},
\ea
\begin{figure}[b]\centering
\includegraphics[width=12cm]{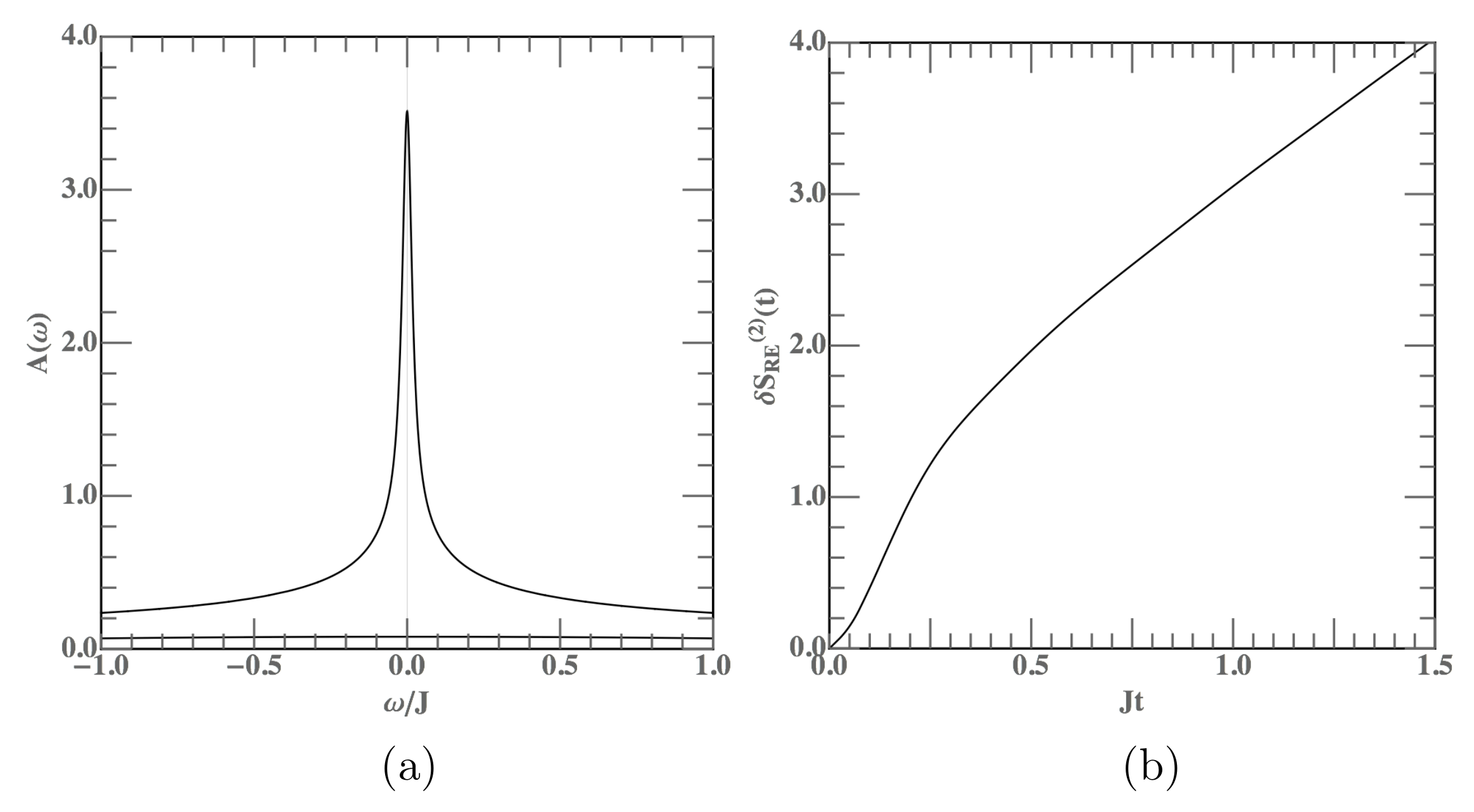}
\caption{Heating of the `` Black Hole ''. The figure shows the R\'{e}nyi entropy response of SYK$_4$ with ${\cal J}=4$ and $\beta=10$ coupling to SYK$_2$ model with ${\cal J}'=8$ and $\beta_{\rm E}=2$. A short time $t^2$ growth and a long time $t$ linear growth are shown. }
\label{Heating}
\end{figure}
%
%
%\ba
%G_{q,R}(\omega)=-ib\left(\frac{\pi}{\beta}\right)^{2\Delta_q-1}2^{2\Delta_q}\cos(\pi\Delta_q)\frac{\Gamma(1-2\Delta_q)\Gamma(-i\beta\omega/2\pi+\Delta_q)}{\Gamma(-i\beta\omega/2\pi+1-\Delta_q)}
%\ea
where $b=(1/4\pi {\cal J}^2)^{1/4}$. The spectral function of the environment is 
\ba
{\cal A}_\xi(\omega)=\sqrt{({\cal J}')^2-\omega^2}.
\ea
According to the entropy response theory
\ba
\delta S_{\rm RE}^{(2)}(t)=\frac{4g^2 NM}{{\rm Tr}(\hat{\rho}_0^2)}\int_0^{t/2} dt_0\int_{-{\cal J'}}^{\cal J'}\!\!\!d\omega'\int_{-{\cal J}}^{\cal J}\!\!\!d\omega\frac{\sin(\omega-\omega')t_0}{\omega-\omega'}{\cal W}_{\beta\beta_E}^F(\omega,\omega'){\cal A}_{\cal O}(\omega){\cal A}_{\xi}(\omega')
\ea
In extremely short time, $\sin(\omega-\omega')t_0\sim (\omega-\omega')t_0$, therefore we have
\ba
\delta S_{\rm RE}^{(2)}(t)=\frac{1}{2}\kappa_F t^2,
\ea
where $\kappa_F=8g^2 NM/{\rm Tr}(\hat{\rho}_0^2)\int_{-{\cal J'}}^{\cal J'}d\omega'\int_{-{\cal J}}^{\cal J}d\omega {\cal W}_{\beta\beta_E}^F(\omega,\omega'){\cal A}_{\cal O}(\omega){\cal A}_{\xi}(\omega')$. 
In longer time, $\sin(\omega-\omega')t_0/(\omega-\omega')=\delta(\omega-\omega')$, therefore

\ba
\delta S_{\rm RE}^{(2)}(t)=\lambda t,
\ea
where $\lambda$ is the entropy rate, whose value is
\ba
\lambda=\frac{4g^2 NM}{{\rm Tr}(\hat{\rho}_0^2)}\int_{-{\rm min}({\cal J'},{\cal J})}^{{\rm min}({\cal J'},{\cal J})}\!\!\!d\omega {\cal W}_{\beta\beta_E}^F(\omega,\omega){\cal A}_{\cal O}(\omega){\cal A}_{\xi}(\omega).
\ea
Here $\min(a,b)$ takes the smaller one in $a$ and $b$. Here $\lambda$ can be negative. The general picture for the entanglement entropy growth is a quadratic growth at early time and then crossover to a linear growth ( or decrease ) at late time. As the late time behavior can be a decrease of entropy, therefore we can see the signal of evaporation in our perturbative calculation even if it is not early time behavior. 

Here we stress that for $\sin(\omega-\omega')t_0/(\omega-\omega')\propto\delta(\omega-\omega')$, we need a really large $t_0$. Indeed this approximation is better satisfied when ${\rm min}({\cal J}',{\cal J})t_0$ is really large. Therefore before that time scale, there are always oscillations in $\delta S^{(2)}_{\rm RE}(t)$, but these oscillations are slow. In long time limit, we find only if $\beta=\beta_{\rm E}$, we have $\delta S^{(2)}_{\rm RE}=0$. However, in some period of time, $\delta S^{(2)}_{\rm RE}\sim0$ for quite remarkable time scale, which may seems like prethermalization. In this section, we only focus on the situation where $\beta$ and $\beta_{\rm E}$ has large difference. 
\begin{figure}[b]\centering
\includegraphics[width=12cm]{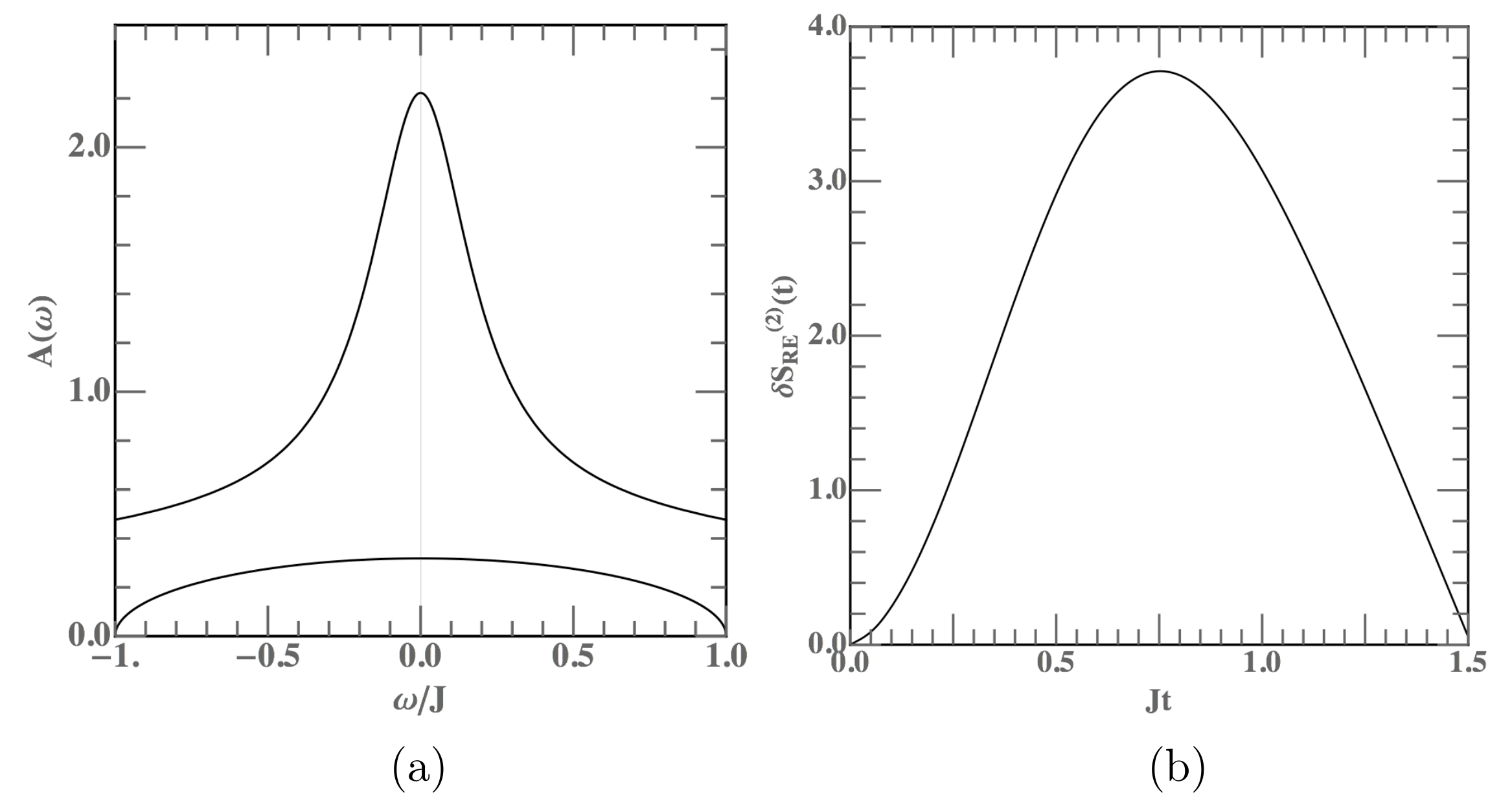}
\caption{Evaporation of "Black Hole", In (a) we show the spectral function of the system ( SYK$_4$ model , $\beta=2$, ${\cal J}=2$ ) and the environment ( SYK$_2$ model with $\beta=20$, ${\cal J}'=2$ ); In (b), we show the R\'{e}nyi entropy response of the sudden coupling between SYK model and the environment. As the environment is at low temperature, it is an evaporation process. We can see a linear growth follows a quadratic growth, then a sharp linear drop follows. The shape of the curve is alike the Page curve in black hole evaporation. $g^2 NM$ is tuned for showing the figures, but these values are not important at this level.}
\label{Evaporation}
\end{figure}

In the following, we present the numerical calculation for mainly two kinds of process. First of all, we discuss the situation of heating where the system is at low temperature and the environment is at high temperature. As is shown in Fig.~\ref{Heating}, we couple a SYK$_4$ model with ${\cal J}=4$, $\beta=10$ to a SYK$_2$ model with ${\cal J}'=8$ and $\beta_{\rm E}=2$. This situation mimic a Black Hole coupling to a hot environment. Just as our analysis in the last section, the R\'{e}nyi entropy experienced a quadratic growth in the short time limit and then crossover to a linear growth in longer time scale.

Second, we study a SYK$_4$ model with ${\cal J}=2$, $\beta=2$ coupling to a SYK$_2$ model with ${\cal J}'=2$ and $\beta_{\rm E}=20$. The environment is at low temperature, hence it describes a `` Black Hole '' evaporation process.  A Page curve alike behavior is expected and we find that even if the method is perturbative, the non-monotonic behavior can be captured as is shown in Fig.~\ref{Evaporation}.

\subsection{Heating and Evaporation with non-Markovian Environment}

%The conformal limit is satisfied when $N\gg \beta{\cal J}\gg1$. When $N$ and $\beta{\cal J}$ are comparable, the low temperature ( $\beta>{\cal J}/N$ ) spectrum of SYK$_4$ model changes from a divergent zero frequency density of states to zero density of states. Here we try to understand how the low energy spectrum property will change the behavior of R\'{e}nyi entropy response.

%At zero temperature, the imaginary time Green's function of SYK$_4$ model can be obtained\cite{Altland16, Altland17, Witten, Verlinde} as,

Here we assume the environment's hamiltonian is
\ba
\hat{H}_{\rm E}=\sum_{\alpha=1}^M\sum_{jk=1}^{N} (J'_{jk,\alpha L}-\mu)\hat{\psi}_{j,\alpha L}^\dag\hat{\psi}_{k,\alpha L}^{}+\sum_{\alpha=1}^M\sum_{jk=1}^{N} (J'_{jk, \alpha R}+\mu)\hat{\psi}_{j,\alpha R}^\dag\hat{\psi}_{k,\alpha R}^{},
\ea
where
\ba
\overline{J'_{jk,\alpha L(R)}}=0,\hspace{4ex}\overline{(J_{jk, \alpha L(R)}^{\prime})^2}=\frac{1}{N}({\cal J}^{\prime})^2.
\ea
\begin{figure}[h]\centering
\includegraphics[width=12cm]{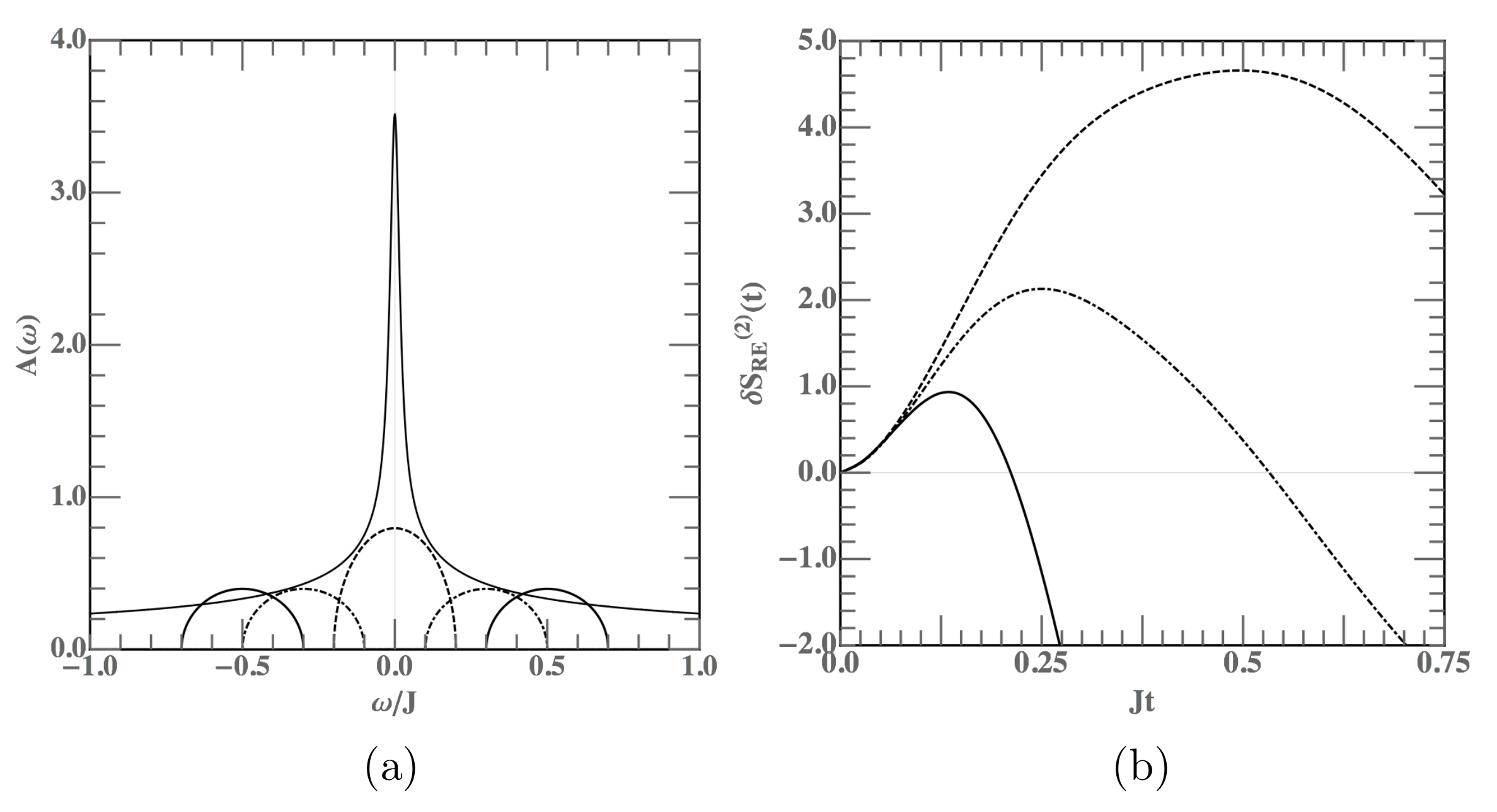}
\caption{Here we take ${\cal J}=4$, $\beta=10$, ${\cal J}'=0.8$. In (a) we give the spectral function of the system and the environment. The dashed line is for $\mu=0.$, dot-dashed line is for $\mu=1.2$ and the solid line is for $\mu=2$. The gap of the spectrum is increasing from $0$ to 1.2. In (b) we show the R\'{e}nyi entropy response for $\beta_{\rm E}=2$, $\mu=0.$ as dashed line; $\mu=1.2$ case as dot-dashed line and $\mu=2.$ as solid line. }\label{Gapped}
\end{figure}
The interactions between the system and the environment is
\ba
\hat{V}=g\sum_{j=1}^N\sum_{\alpha=1}^M\sum_{\sigma=L,R} (\hat{\chi}_j\hat{\psi}^\dag_{j,\alpha\sigma}+\hat{\chi}_j\hat{\psi}^{}_{j,\alpha\sigma})
\ea
For our problem, this model change is equivalent to a simple change in environment's spectral function.
\ba
{\cal A}_\xi(\omega)=\frac{1}{\pi {\cal J'}^2}\left(\sqrt{{\cal J}^{\prime, 2}-(\omega-\mu)^2}+\sqrt{{\cal J}^{\prime, 2}-(\omega+\mu)^2}\right)
\ea
Here we assume $\mu>{\cal J}'$, such that we have a gapped system. Here we take ${\cal J}'=0.5$, $\mu=1$. Again we apply Eq.~(\ref{REBOSE}), then we show the numerical results in Fig.~\ref{Gapped} . We compared the results with the same parameter ${\cal J}'$ with previous spectral function. For a case where $\beta_{\rm E}<\beta$, the presence of a gap can speed up the entanglement entropy decreasing.

\begin{figure}[h]\centering
%\includegraphics[width=11.5cm]{RERESOS_M.pdf}\\
%\hspace{0.7cm}(a)\\
%\includegraphics[width=12cm]{RERESOS.pdf}\\
%\hspace{0.7cm}(b)
\includegraphics[width=11.cm]{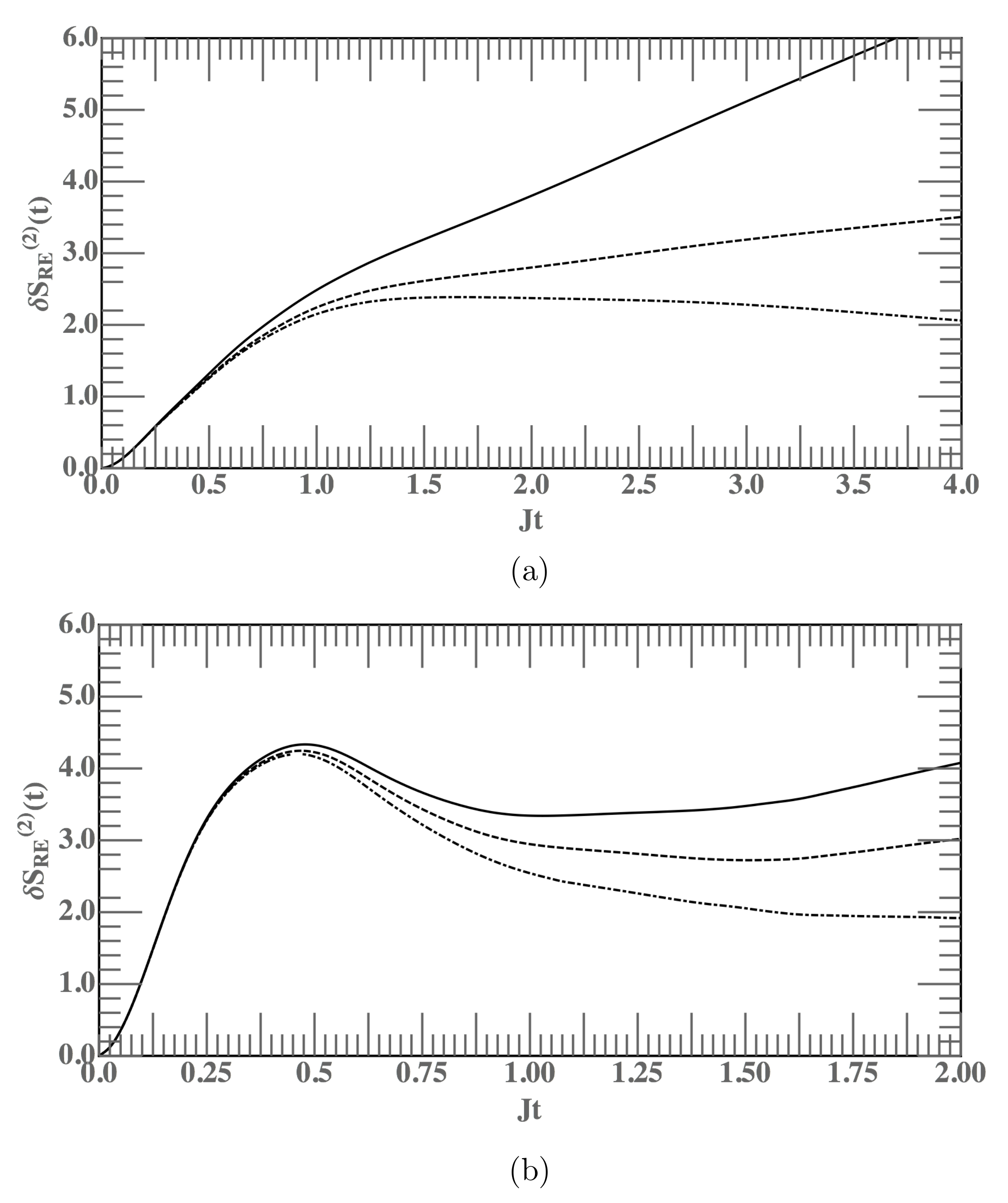}
\caption{In (a) we show the RE response for environment with gapless spectrum, $\beta=10$, ${\cal J}=4$, $\beta_{\rm E}=4.5$, 8 and 12 respectively with solid line, dashed line and dot-dashed line, ${\cal J}'=0.8$; In (b) we show an oscillation behavior of $\delta S_{\rm RE}^{(2)}$ when $\beta_{\rm E}\approx\beta$. Here $\beta$ is fixed as $10$. Here the solid line has a temperature $\beta_{\rm E}=4.5$. The dashed line is at the temperature $\beta_{\rm E}=8$ and the dot-dashed line is at the environment temperature $\beta_{\rm E}=12$. }
\end{figure}
A very non-trivial and interesting dynamical behavior is observed that the entropy experiences an oscillation before a late time linear growth. This situation happens when the environment temperature is close to the system's temperature. We find the gap in the spectrum can result non-trivial memory effect in entropy dynamics where the heat flow direction can change at least twice. Deeper reasons for the extra oscillation behavior in RE dynamics is remain to be  further understood. By comparing the behavior of the gapless spectrum and the gapped spectrum under similar circumstance, we find the gapped spectrum bring the oscillation of RE to shorter time scale.

\section{Conclusion}

In summary, we construct a perturbative method for calculating the R\'{e}nyi entropy for a sudden coupling between a system and an environment. On condition that the coupling between two systems are linear coupling, we obtained general formula for R\'{e}nyi entropy response for bosonic systems and fermionic systems respectively. One can find the entropy response is related to the spectral function of the system and the environment as well as a distribution kernel function ${\cal W}^{B,F}$. We find the short time behavior of R\'{e}nyi entropy response follows a $t^2$ law, and later it follows a $t$ linear law. Fermi statistics can result a non-monotonic behavior in R\'{e}nyi entropy dynamics which may have an gravitational explanation as a Page curve in black evaporation. Further, among two reasons for a bath being non-Markovian, we find the distribution kernel plays the vital role in the behavior of entanglement entropy dynamics, while the spetral function of environment controls the detail behavior of the entropy dynamics. Further understanding of these perturbative results in gravity language, as well as the back action of the environment to the sudden coupling of the system are left for further study.

At the closing stage of this paper, we notice a similar paper on this topic as arXiv: 2011.09622 by Dadras and Kitaev\cite{Kitaev20}. 

\appendix

%\section{The Spectrum Decomposition Theorem for Green's functions}

\acknowledgments

We thank Pengfei Zhang, Yiming Chen and Hui Zhai for discussions. This work is supported by NSFC under Grant No. 11734010, Beijing Natural Science Foundation (Z180013).

%\paragraph{Note added.} 


\end{document}